\documentclass[journal]{IEEEtran}
\usepackage[edges]{forest}
\usetikzlibrary{arrows.meta}
\usepackage{graphicx,times,amsmath,cite,enumerate}
\usepackage{kantlipsum}
\usepackage{threeparttable}
\usepackage{epsfig}
\usepackage{amssymb}
\usepackage{latexsym}
\usepackage{psfrag}
\usepackage{setspace}
\usepackage{color}
\usepackage{verbatim}
\usepackage{amsthm}
\usepackage{footnote}
\usepackage{textcase}
\usepackage{array}
\newcolumntype{P}[1]{>{\centering\arraybackslash}p{#1}}
\usepackage{float} 
\usepackage{multirow}
\usepackage{textcomp}
\usepackage{forest}
\PassOptionsToPackage{hyphens}{url}
\usepackage{url}
\usepackage{physics}
\usepackage{hyperref}
\usepackage{etoolbox}
\usepackage[edges]{forest}
\usepackage[ruled,vlined]{algorithm2e}
\usepackage{siunitx}

\usepackage{subfigure}
\usepackage{physics}
\usepackage{tikz}
\usetikzlibrary{automata, positioning}
\usepackage{nomencl}
\usepackage{float}
\usepackage{tabularx,booktabs}
\usepackage{footnote}

\renewcommand{\baselinestretch}{0.98}

\title{\vspace{-0pt} Microgrid Building Blocks for Dynamic Decoupling and Black Start Applications \vspace{-0pt}}

\IEEEoverridecommandlockouts

\begin{document}
\bstctlcite{IEEEexample:BSTcontrol} 


 \author{
    \IEEEauthorblockN{ Samrat Acharya}\IEEEauthorrefmark{1}
    , \IEEEauthorblockN{ Priya Mana}\IEEEauthorrefmark{1}%
    , \IEEEauthorblockN{ Hisham Mahmood}
        , \IEEEauthorblockN{ Francis Tuffner}
        , \IEEEauthorblockN{ Alok Kumar Bharati}
    \\
    \IEEEauthorblockA{\textit{Pacific Northwest National Laboratory (PNNL), Richland, WA, USA}}\\

    \IEEEauthorblockA{(samrat.acharya, priya.mana, hisham.mahmood, francis.tuffner,  ak.bharati)@pnnl.gov \vspace{-8mm}}
}
\maketitle

\def\thefootnote{*}\footnotetext{Samrat Acharya and Priya Mana are joint first authors. 

The Pacific Northwest National Laboratory is operated by the Battelle Memorial Institute for the U.S. Department of Energy under contact DE-AC05-76RL01830.}
\def\thefootnote{\arabic{footnote}}

\begin{abstract}
Microgrids offer increased self-reliance and resilience at the grid's edge. They promote a significant transition to decentralized and renewable energy production by optimizing the utilization of local renewable sources. However, to maintain stable operations under all conditions and harness microgrids' full economic and technological potential, it is essential to integrate with the bulk grid and neighboring microgrids seamlessly. In this paper, we explore the capabilities of Back-to-Back (BTB) converters as a pivotal technology for interfacing microgrids, hybrid AC/DC grids, and bulk grids, by leveraging a comprehensive phasor-domain model integrated into GridLAB-D. The phasor-domain model is computationally efficient for simulating BTB with bulk grids and networked microgrids. We showcase the versatility of BTB converters (an integrated Microgrid Building Block) by configuring a two-microgrid network from a modified IEEE 13-node distribution system. These microgrids are equipped with diesel generators, photovoltaic units, and Battery Energy Storage Systems (BESS). The simulation studies are focused on use cases demonstrating dynamic decoupling and controlled support that a microgrid can provide via a BTB converter.

\end{abstract}
%
\IEEEpeerreviewmaketitle

\section{Introduction}

  The potential for microgrids to provide uninterrupted power supply to communities and commercial establishments is promising. For large scale implementation of microgrids, feeders can be connected through a Back-to-Back (BTB) AC-DC-AC converter not only to enable control of the microgrids, but also to easily re-synchronize with the bulk grid. Liu \textit{et al.} discuss the Microgrid Building Block (MBB) concept in~\cite{chenching_ieeeaccess}, presenting applications of BTB converters beyond the traditional power transfer capability. The applications of the BTB in \cite{chenching_ieeeaccess} include the use of BTB converters for providing secondary control in the microgrid, providing the ability to black start the microgrid, and using the microgrid to provide cranking power by picking up loads in the bulk grid. These BTB applications in \cite{chenching_ieeeaccess} are proof of concepts for MBB based microgrids and are aligned with the DOE-Office of Electricity's focus on modular microgrids.

  Previous work in the area of autonomous power transfer between microgrid and grid interfaces discusses a power-sharing unit composed of three BTB converters at the Point of Common Coupling (PCC)~\cite{giriv2009}. The three BTBs, one for each individual phase, help balance power and coordinate with single-phase distributed generators in the system. Sun \textit{et al.}~\cite{gurrero2015} models a BTB converter at the interconnection of two islanded microgrid. The BTB converter can use local measurements to decide how much power should be exchanged, to provide frequency support to both microgrids~\cite{blaabjergslectedpubs2021}. Bilakanti \textit{et al.}~\cite{bilakanti2019} discuss an Island Interconnection Device (IID) that holds the BTB converter at the PCC of the microgrid responsible for interconnection standards and not the downstream distributed energy resources (DERs). Smith \textit{et al.}~\cite{gm2017b2b} use a simplified system with a single load representing microgrid and an infinite bus representing the grid to propose passivity-based controls for damping oscillations at the PCC. Nutkani \textit{et al.}~\cite{nutkani2013power} study a microgrid with a lumped-load where the BTB converter controls are discussed for power exchange. The study in \cite{nutkani2013power} does not model microgrids in detail, which can limit the understanding of varying impedance due to different topologies and generation mixes. Other applications of medium voltage BTB converters highlight their utility to manage and control power transfer between DER-rich feeders by interconnecting two distribution feeders from an economics standpoint~\cite{nrel2023}. A commercial product called Grid Block that connects and provides decoupling for an individual load with the rest of the grid is in a pilot state of implementation~\cite{grid_block}.

  While the previous research on BTB converters discuss the control and operation of the converters with lumped-loads, there is a lack of system-level studies to understand BTB converter applications for changing power system topology, networking microgrids, and efficient converter operation in the presence of variable DERs such as PV, BESS and diesel generators. The BTB converter interface at the PCC for a microgrid is able to efficiently transfer power in networked microgrid mode, or when connected to the bulk grid. One limiting factor for this is the lack of phasor domain dynamic model that can simulate large systems over a longer time duration without compromising on the dynamic details during a transient event. To enable this, we developed a phasor-domain model of a back-to-back converter while emulating the electro-mechanical grid dynamics~\cite{mahmood2024dynamic}. In this paper, our main contribution is the simulation of realistic use case scenarios for electrically connected but decoupled networked microgrids using a BTB converter as a part of an MBB to enable scalable system-level phasor-domain studies. The typical use cases of the MBB are discussed next followed by the implementation of the BTB converter model in Section~\ref{sec:BTB_phasor_domain} and a case study of networked microgrid use-cases in Section~\ref{sec:case_study} followed by the conclusion in Section~\ref{sec:conclusion}. 

   \begin{table}[H]
\centering
\begin{tabular}{ |p{0.46\textwidth}| }
 \hline
\rule{0pt}{2.5ex}  This paper is accepted for publication in IEEE PES Grid Edge Technologies Conference \& Exposition 2025, San Diego, CA. The complete copyright version will be available on IEEE Xplore when the conference proceedings are published. \\
 \hline
\end{tabular}
\end{table}

\section{BTB Converter as A key MBB Component - Modeling and Use Cases}
\label{sec:BTB_phasor_domain}
The key uniqueness of MBB-based microgrids is the ability for modularization and standardized design of the microgrid components. A key component enabling such modularization is the BTB converter at the microgrid PCC. With BTB converter added to microgrids, the power transfer beyond the microgrid can be controlled ensuring the stable operation of the microgrid while being electrically connected to the neighboring system, either the bulk system or other microgrids. 
\vspace{-4mm}
\subsection{Typical Use Cases of MBB}
The typical use cases of MBB-based microgrids are below.

\begin{enumerate}
    \item  Microgrids with limited energy sources mainly designed to meet the microgrid peak loads. During normal operations at low loading, the microgrid may have reserve resources and can choose to dispatch those for ancillary services or grid support beyond the microgrid.

    \item Microgrids that are designed to mainly operate and sustain critical loads that need to be dynamically decoupled from the rest of the system to ensure disturbances are isolated from the MBB based microgrid.
    
    \item Microgrids that can be developed with minimal lead time owing to standardized designs. As a part of the MBB, a BTB converter at the PCC with available DERs enables such a possibility. 
    
\end{enumerate}

The common aspect in these typical use cases is the ability of the microgrid to operate and supply a set of critical loads with limited resources by ensuring safe and reliable operation. This is while the system experiences possible large and small disturbances such as switching of medium to large loads, generator switching, and intermittence in renewables. 



\begin{figure}[!b]
    \centering
    \vspace{-5pt}
\includegraphics[width=0.95\columnwidth, clip=true, trim= 0mm 0.8mm 0mm 0mm]{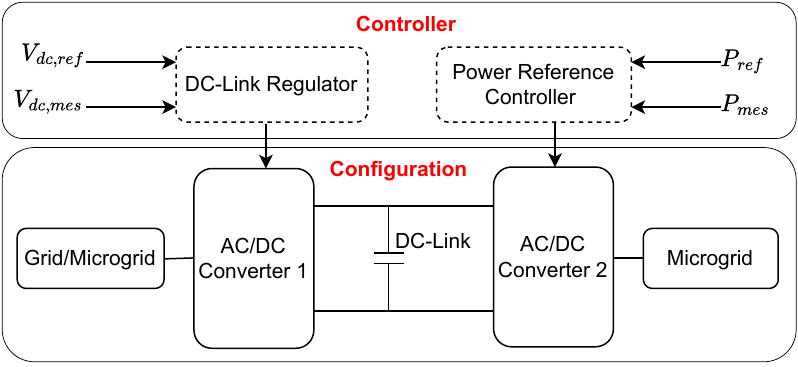}
    \caption{High-level architecture of BTB converter configuration and control.}
    \label{fig:b2b_block_diagram}
\end{figure}

\vspace{-4mm}
\subsection{Implemntation of Phasor Domain Model of BTB Converter}
\label{subsec:phasor_domain_model}
\begin{figure}[!t]
    \centering
     \vspace{-1mm}
\includegraphics[width=0.99\columnwidth, clip=true, trim= 1mm 1.1mm 2.8mm 0mm]{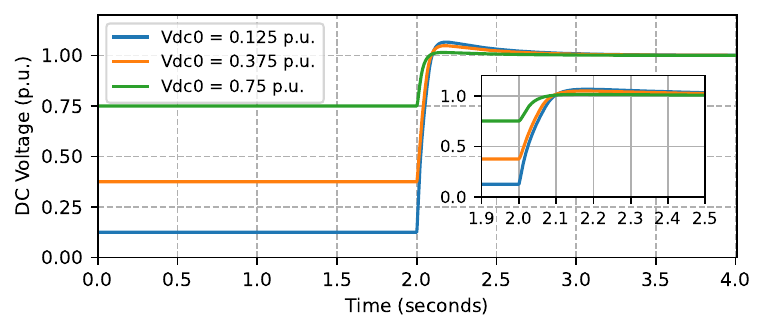}
    \caption{DC-link voltage building dynamics of the Back-to-Back converter.}
    \label{fig:btb_Vdc_build_up}
    \vspace{-5mm}
\end{figure}

We use the phasor domain model of the BTB converter, detailed in \cite{mahmood2024dynamic}, to study system-level use cases for an MBB.  As Fig.~\ref{fig:b2b_block_diagram} shows, the BTB converter has two AC/DC converters connected together by DC-link. 
Among these two converters, one is responsible for regulating the DC-link voltage of the BTB converter (AC/DC Converter 1), while the other is responsible for controlling the power flowing through the BTB converter (AC/DC Converter 2). The AC/DC Converter 1 is modeled as grid-following current source converter and the AC/DC Converter 2 can operate in either grid forming or grid following mode. The model captures the charging and discharging dynamics of the DC-link, which is an important aspect of the BTB converter model. Fig.~\ref{fig:btb_Vdc_build_up} shows the DC-link voltage build-up dynamics for different initial values ($Vdc_0$). While this phasor domain model of the BTB converter can be implemented in many simulation tools, we use GridLAB-D for the modeling the MBB based microgrids and simulating the use cases. GridLAB-D is an open-source simulation platform developed by the Pacific Northwest National Laboratory for modeling and simulating the operation and behavior of three-phase unbalanced distribution power systems in steady state, quasi steady state, and dynamic timescales \cite{8307765,8267335}.



\vspace{-1mm}
\section{Case Study}
\label{sec:case_study}

\begin{figure}[!b]
    \centering
    \vspace{-10pt}
\includegraphics[width=0.95\columnwidth, clip=true, trim= 0mm 1.1mm 0mm 0mm]{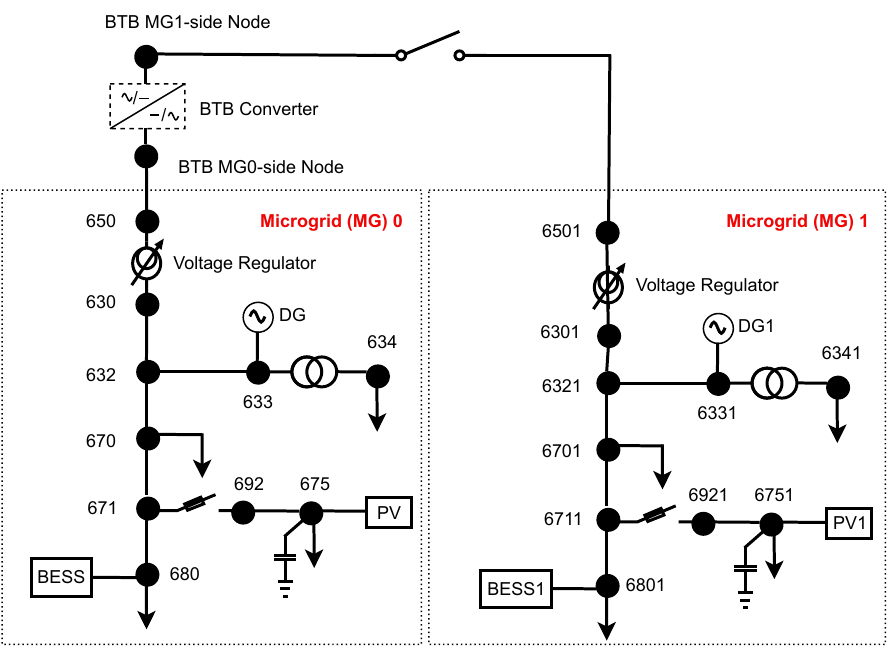}
    \caption{Two microgrids connected via a Back-to-Back converter.}
    \label{fig:two_MGs_system}
\end{figure}


We use two identical microgrids connected via a BTB converter as a test system, shown in Fig.~\ref{fig:two_MGs_system}. Both microgrids (MG0 and MG1) are modified IEEE 13-node systems. Given the identical configuration of MG0 and MG1, we only discuss MG0 components further. The system-level voltage is 4.16 kV except for node 634, which is stepped down by a 1 MVA, 4.16 kV/480 V transformer to 480 V. The microgrid consists of a  BESS at node 680, a 1.6 MW PV unit at node 675, and a 3 MVA diesel generator at node 633. The BESS capacity varies across the simulation cases below. The MG0 microgrid is connected to another microgrid through the BTB converter upstream of node 650. The three-phase constant power balanced microgrid loads are distributed at nodes 634 ($400$ kW and $290$ kVAR), 670 ($200$ kW and $116$ kVAR), 675 ($843$ kW and $462$ kVAR) and 680 ($1155$ kW and $660$ kVAR). A shunt capacitor of 86.52 kVAR capacity is installed at node 675 to help boost the node voltage. A voltage regulator between nodes 650 and 630 regulates the voltage based on the voltage measured at node 680.  Both of the AC/DC converters in the BTB converter are rated as 3~MVA with the nominal DC-link voltage set to 8000~V. In this paper, we assume MG0 is the MBB-based microgrid with excess resources (parent microgrid) and MG1 as microgrid with limited resource.

We conducted the case study in GridLAB-D on a Windows laptop with an i7-11850H processor and 16 GB RAM. All the simulations below are completed in under 1 minute. This shows the efficient computation capability of the developed phasor model which facilitates for the faster simulation of high-density power networks.




\begin{figure}[!b]
\centering

\subfigure[\label{fig:mg1_load_flexible_p}]{\includegraphics[width=0.92\columnwidth, clip=true, trim= 2.5mm 7mm 2.5mm 2.5mm]{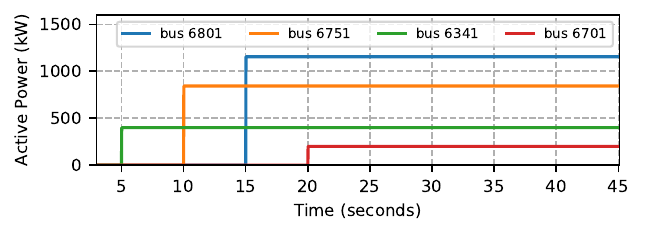}}

\subfigure[\label{fig:mg1_bess_flexible_p}]{
\includegraphics[width=0.92\columnwidth, clip=true, trim= 2.5mm 7mm 2.5mm 2.5mm]{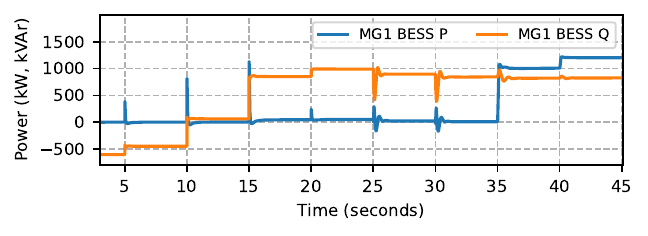}}

\subfigure[\label{fig:mg1_pv_flexible_p}]{\includegraphics[width=0.92\columnwidth, clip=true, trim= 2.5mm 2mm 2.5mm 2.5mm]{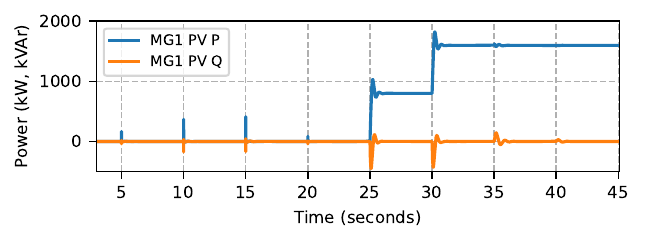}}

\vspace{-3mm}
\caption{MG1 - the networked microgrid: (a) Loads in MG1 are picked up with support from MG0 via BTB converter, (b) MG1 BESS operation to avoid overloading, and (c) PV at MG1 is brought up once the microgrid is stabilized.}
\vspace{-5mm}
\label{fig:mg1_flexible_P}
\end{figure}

\begin{figure}[!htb]
\centering
\subfigure[\label{fig:mg0_bess_flexible_p}]{
\includegraphics[width=0.92\columnwidth, clip=true, trim= 2.5mm 7mm 2.5mm 2.5mm]{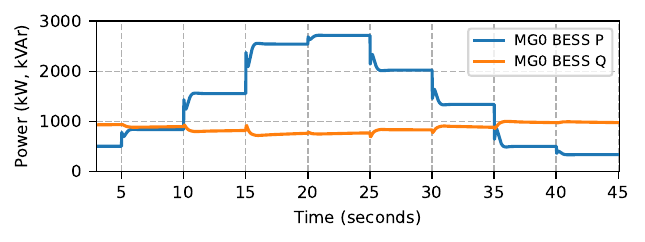}}

\subfigure[\label{fig:mg0_dg_flexible_p}]{\includegraphics[width=0.92\columnwidth, clip=true, trim= 2.5mm 7mm 2.5mm 2.5mm]{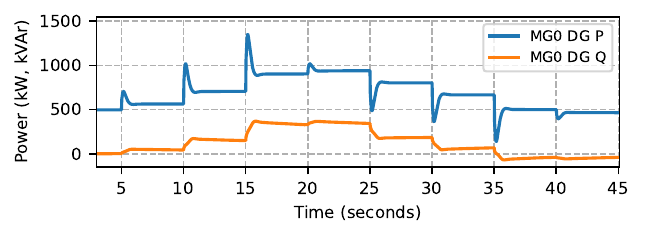}}

\subfigure[\label{fig:mg0_pv_flexible_p}]{\includegraphics[width=0.92\columnwidth, clip=true, trim= 2.5mm 2mm 2.5mm 2.5mm]{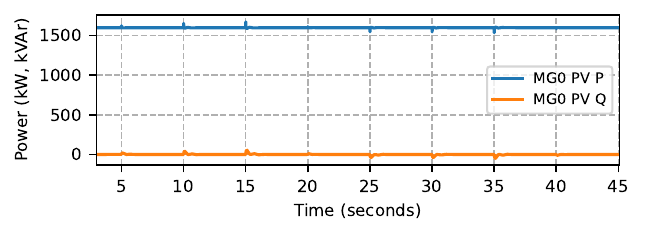}}

\vspace{-3mm}
\caption{MG0 - the parent microgrid: Changes in (a) BESS, (b) DG and (c)~PV output in MG0 to support MG1 for flexible power exchange capability.}
\label{fig:mg0_flexible_P}
\end{figure}

\begin{figure}[!htb]
\centering
\subfigure[\label{fig:btb_flexible_P}]{
\includegraphics[width=0.92\columnwidth, clip=true, trim= 2.5mm 7mm 2.5mm 2.5mm]{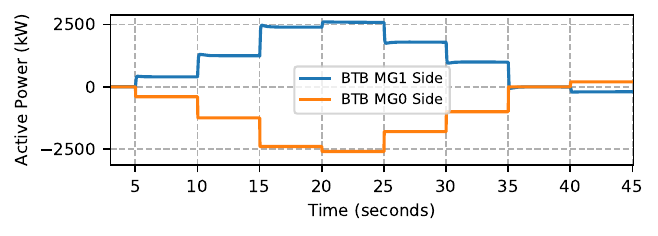}}

\subfigure[\label{fig:btb_Vdc_flexible_P}]{\includegraphics[width=0.92\columnwidth, clip=true, trim= 2.5mm 7mm 2.5mm 2.5mm]{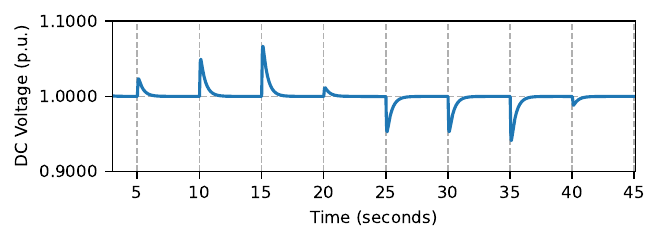}}

\subfigure[\label{fig:freq_flexible_P}]{\includegraphics[width=0.92\columnwidth, clip=true, trim= 2.5mm 2mm 2.5mm 2.5mm]{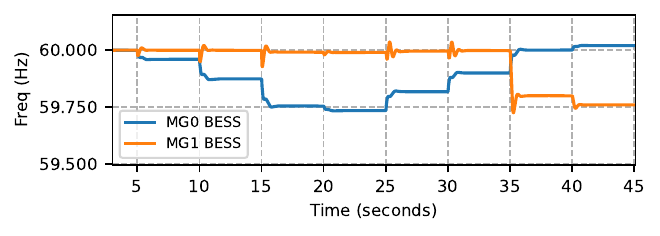}}
\vspace{-2mm}
\caption{(a) Power exchange through the BTB converter, (b) BTB converter DC-link voltage, and (c) frequencies at MG0 and MG1 during power exchange.}
\label{fig:btb_flexible_PVdcf}
\vspace{-4mm}
\end{figure}
\vspace{-4mm}
\subsection{Flexible Power Exchange Between Microgrids}
In this section, we demonstrate the capability of the BTB converter that allows grid operators for flexible power exchange between the two microgrids in Fig.~\ref{fig:two_MGs_system}. Grid/microgrid operators use this flexible power exchange capability to leverage low-cost generation and/or support each other during an emergency. To demonstrate this use case, we start with a scenario where all the loads in MG0 are online (energized), and all loads in  MG1 are offline (de-energized). In MG0, the DG, PV, and BESS are active. In MG1, the PV and DG are disconnected to create the limited resources case at the start of the simulation. As the simulation progresses, the loads in MG0 are energized at bus 6341, bus 6751, bus 6801, and bus 6701 at times $t=5$, $t=10$, $t=15$, and $t=20$ seconds as shown in Fig.~\ref{fig:mg1_load_flexible_p}. To avoid overloading the BESS at MG1, we inject the required power by the MG1 from MG0 via the BTB converter.
 Once the load support is requested by the BTB converter from MG0 at $t=5$ seconds, the BESS and DG in MG0, operating as grid-forming inverter with $1\%$ P-f droop, and at $3\%$ P-f droop respectively, increase their power generation and share the load as shown in Fig.~\ref{fig:mg0_bess_flexible_p} and~\ref{fig:mg0_dg_flexible_p}. PV at MG0 is operating at its Maximum Power Point Tracking (MPPT) capacity at $1600$ kW as shown in Fig.~\ref{fig:mg0_pv_flexible_p}.
 
After all the loads in MG1 are online, PV at MG1 is taken into service at $t=25$ seconds with an initial power reference of $800$~kW to be ramped to MPPT capacity ($1600$ kW) at $t=30$~seconds as shown in Fig.~\ref{fig:mg1_pv_flexible_p}. Now the PV and the BESS in MG1 microgrid that is networked with the MG0 are able to fully support the local loads and the power exchange from MG0 to MG1 is no longer required. To demonstrate the bi-directionality of power flow via the BTB converter, the excessive supply of power form MG1 is sent back to MG0 at $t=40$ seconds as shown in Fig.~\ref{fig:btb_flexible_P}. During these power exchange events, BTB converter's DC-link voltage is maintained between 1.068 per unit ($8550$ V) and 
0.94 p.u. ($7529$ V) as shown in Fig.~
\ref{fig:btb_Vdc_flexible_P}. The frequency at both MG0 and MG1 measured at their terminal of grid-forming BESS are shown in Fig.~\ref{fig:freq_flexible_P}.
This use case demonstrates the successful exchange of power from a MG0 to a MG1 and vice-versa via the BTB converter while the frequency at each microgrid is regulated independently.

\vspace{-4mm}
\subsection{Dynamic Decoupling of Microgrids}
\label{sec:dynamic_decoupling}
\begin{figure}[!htb]
\centering
\subfigure[\label{fig:btb_p_dynamic_decoupling}]{
\includegraphics[width=0.92\columnwidth, clip=true, trim= 2.5mm 7mm 2.5mm 2.5mm]
{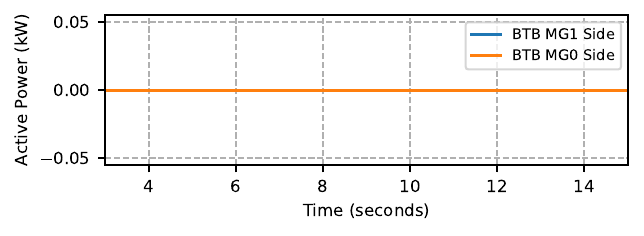}}

\subfigure[\label{fig:btb_Vdc_dynamic_decoupling}]{\includegraphics[width=0.92\columnwidth, clip=true, trim= 2.5mm 7mm 2.5mm 2.5mm]{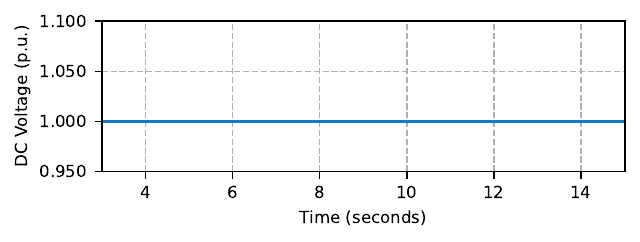}}

\subfigure[\label{fig:freq_dynamic_decoupling}]{\includegraphics[width=0.92\columnwidth, clip=true, trim= 2.5mm 2mm 2.5mm 2.5mm]{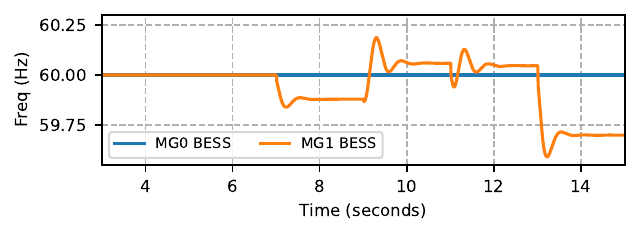}}

\vspace{-2.5mm}
\caption{(a) No power is exchanged through the BTB converter, (b) DC-link voltage of BTB converter is maintained at a stable value, and (c) the frequencies at MG0 is kept constant even as MG1 frequency varies - an example of dynamic decoupling capability.}
\label{fig:btb_dynamic_decoupling}
\end{figure}

\begin{figure}[!htb]
\centering
\subfigure[\label{fig:mg1_load_dynamic_decoupling}]{\includegraphics[width=0.92\columnwidth, clip=true, trim= 2.5mm 7mm 2.5mm 2.5mm]{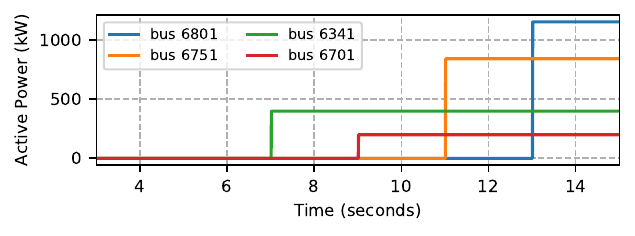}}

\subfigure[\label{fig:mg1_bess_dynamic_decoupling}]{
\includegraphics[width=0.92\columnwidth, clip=true, trim= 2.5mm 7mm 2.5mm 2.5mm]{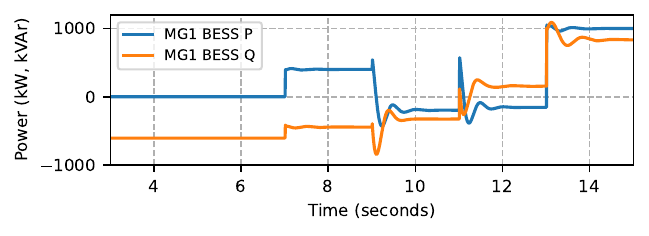}}

\subfigure[\label{fig:mg1_pv_dynamic_decoupling}]{\includegraphics[width=0.92\columnwidth, clip=true, trim= 2.5mm 2mm 2.5mm 2.5mm]{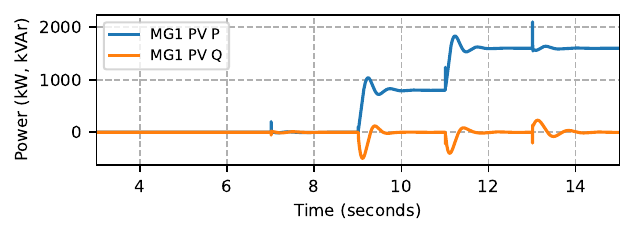}}

\vspace{-2.5mm}
\caption{MG1 networked microgrid: (a) Loads in MG1 are picked up using local generation in MG1. Increase in power supply by MG1 (b) BESS, and (c) PV to support local loads in MG1.}
\vspace{-5mm}
\label{fig:mg1_dynamic_decoupling}
\end{figure}

This section demonstrates the capability of the BTB converter to seamlessly network microgrids and decouple the dynamics of one microgrid from the other. Grid/microgrid operators need this capability to limit the cascading of disturbances into a healthy system while remaining networked. To demonstrate this capability, we start the simulation with a normally operating MG0 in islanded mode where all the loads and DERs are online. On the other hand, MG1 is at an OFF state, where all the loads and generation are offline. 
Since the MG1 is disconnected from the BTB converter (via an open switch between MG1-side BTB node and node 6501), we use a $0.1$ kW inverter as a grid forming source connected to the MG1-side BTB converter to build the BTB DC-link voltage. At $t=4$ seconds, we connect MG1 to MG0 by closing the switch between the BTB MG1-side node and node 6501, shown in Fig.~\ref{fig:two_MGs_system}. As Figs.~\ref{fig:btb_p_dynamic_decoupling},~\ref{fig:btb_Vdc_dynamic_decoupling} respectively show: there is no power exchange between the microgrids via BTB converter after MG0 is connected with MG1; DC-link voltage of the BTB converter is stable at $1$ p.u. 
($8000$ V) during the switching and throughout the simulation. At MG1, at $t=7$, $t=9$, $t=11$, and $t=13$  seconds, the loads at bus 6341, bus 6701, bus 6751, and bus 6801 are energized respectively, as shown in Fig.~\ref{fig:mg1_load_dynamic_decoupling}. To supply these loads, MG1 BESS increases the power output at $t=7$ second as shown in Fig.~\ref{fig:mg1_bess_dynamic_decoupling}. PV starts to supply power at $t=9$ seconds ($800$ KVA-- 50\% of its rating) and operates at MPPT at $t=11$ seconds. All these load increments at MG1 are handled by its PV and BESS, with no dynamic interaction with the parent microgrid MG0 despite their physical connection.

The dynamic decoupling capability enables grids with different frequencies to operate synergistically. As Fig.~\ref{fig:freq_dynamic_decoupling} shows, both grid forming BESS converters at MG0 and MG1 set their frequencies according to their $1\%$ P-f droop control in their respective microgrids.

\begin{figure}[!t]
\centering
\vspace{-5mm}
\subfigure[\label{fig:btb_P_blackstart}]{\includegraphics[width=0.92\columnwidth, clip=true, trim= 2.5mm 7mm 2.5mm 2.5mm]{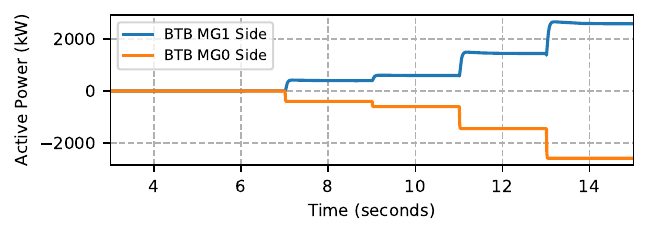}}

\subfigure[\label{fig:btb_Vdc_blackstart}]{
\includegraphics[width=0.92\columnwidth, clip=true, trim= 2.5mm 7mm 2.5mm 2.5mm]{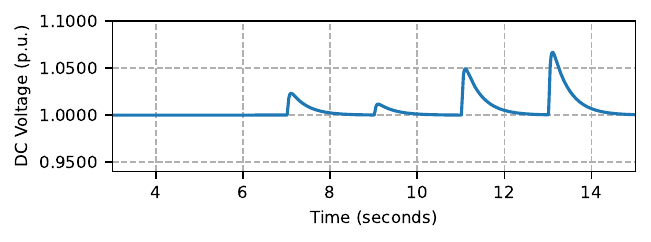}}

\subfigure[\label{fig:freq_blackstart}]{\includegraphics[width=0.92\columnwidth, clip=true, trim= 2.5mm 2mm 2.5mm 2.5mm]{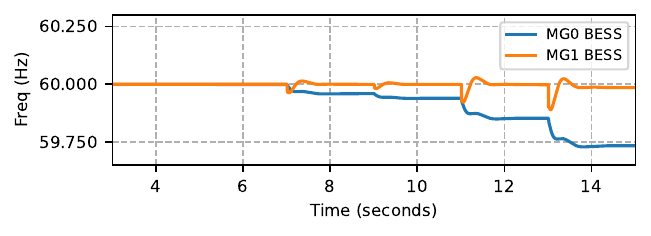}}
\label{fig:mg1_blackstart}

\caption{(a) BTB converter power, (b) DC-link voltage of BTB converter, and (c) frequencies at MG0 and MG1 for black start capability of BTB converter.}
\vspace{-3.5mm}
\end{figure}

\vspace{-4mm}
\subsection {Black Start of Microgrids}
\label{subsec:blackstart}
\begin{figure}[!t]
\centering
\vspace{-5mm}
\subfigure[\label{fig:mg0_bess_blackstart}]{\includegraphics[width=0.92\columnwidth, clip=true, trim= 2.5mm 7mm 2.5mm 2.5mm]{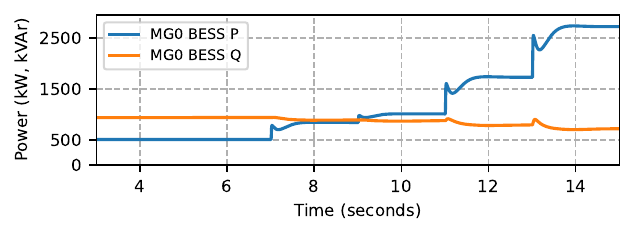}}
\subfigure[\label{fig:mg0_dg_blackstart}]{
\includegraphics[width=0.92\columnwidth, clip=true, trim= 2.5mm 7mm 2.5mm 2.5mm]{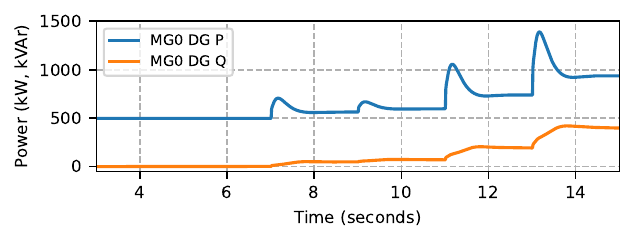}}
\subfigure[\label{fig:mg1_bess_decoupling}]{\includegraphics[width=0.92\columnwidth, clip=true, trim= 2.5mm 2mm 2.5mm 2.5mm]{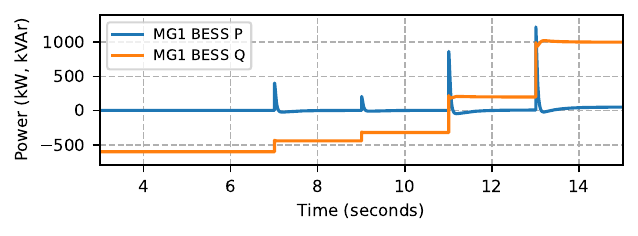}}

\vspace{-1mm}
\caption{Power outputs for black start capability of BTB converter; (a) MG0 BESS, (b) MG0 DG, (c) MG1 BESS.}
\label{fig:mg0_blackstart}
\vspace{-3mm}
\end{figure}

This section demonstrates the capability of the BTB converter to black start MG1 using power generation at MG0. Grid/microgrid operators need this capability to help energize parts of the microgrids that do not have generator availability but slowly pick up (e.g., PV-rich microgrids needing black start at night hours or BESS with low state of charge). To demonstrate this capability, we start the simulation with the same operating conditions as in Section.~\ref{sec:dynamic_decoupling}, i.e., MG1 and MG0 both islanded, MG0 loads and DERs energized, and MG1 completely de-energized. At $t=4$ seconds, we network MG0 and MG1 by closing a switch between the BTB MG1-side node and node 6501, shown in Fig.~\ref{fig:two_MGs_system}. After the switch is closed, MG0 picks up the loads at MG1 at $t=7$, $t=9$, $t=11$, and $t=13$ seconds as shown in Fig.~\ref{fig:mg1_load_dynamic_decoupling}. Unlike in sub-section~\ref{sec:dynamic_decoupling}, we will explore energizing the loads of MG1 through the resources in MG0 to create a black start use case. We supply the MG1 loads from MG0 by providing power references to the BTB converter controller at MG0 side. Fig.~\ref{fig:btb_P_blackstart} shows the active power supplied by MG0 to MG1 via the BTB converter. The increase in power is equal to the load brought online at MG1 at the respective times. Fig.~\ref{fig:btb_Vdc_blackstart} shows the BTB converter DC-link voltage, which is less than 1.1 per unit. Fig.~\ref{fig:mg0_blackstart} shows the power supplied by BESS and DG at MG0 and BESS at MG1. Till $t=7$ seconds (when the load at MG1 starts to come online), BESS, DG , and PV at MG0 supply the rated load (2598 kW and 1528 kVAr). However, after $t=7$ seconds, BESS and DG increase their power supply according to their droop constant to meet the MG1 load. During these events, generation at MG1 are assumed to be unavailable. However, the BESS converter at MG1 supplies reactive power in MG1 as shown in Fig.~\ref{fig:mg1_bess_decoupling}.

\vspace{-2mm}
\section{Conclusion}
\label{sec:conclusion}
  While the integration of BTB converters with renewable energy resources has been discussed in previous literature, the use of BTB converter at the microgrid PCC provides a large set of new capabilities to control the microgrid autonomously, such as dynamically decoupling the microgrid frequency from the grid frequency, enabling networked microgrids, and blackstart capability at the grid's edge. This study provides use cases of BTB converters with detailed distribution system modeling combined with simulation of electro-mechanical dynamics of the grid components. The quasi-static nature of the grid simulation allows for long-term analysis, while the electro-mechanical mode captures details of the dominant dynamics of power electronic components such as BTB converter, BESS, and PV inverters, and also diesel generators in the system. This paper presents a feasibility study for a variety of networked microgrid scenarios. It shows how a microgrid can balance itself automatically when a switch is closed to connect two neighboring microgrids using a BTB converter. The dynamic decoupling allows microgrids with different frequencies to be connected via a BTB converter to maintain stable operations in the parent microgrid (MG0). Since the BTB converter is built in the phasor domain as a part of the modularized MBB concept, a 26-node simulation could be completed in less than a minute on a standard laptop. This enables the study to be further expanded to larger systems with hundreds and thousands of nodes to study the practicality of using BTB converters for large networks of microgrids. 
 \vspace{-3mm}
\section*{Acknowledgment}
The authors wish to thank Dan Ton with the U.S. Department of Energy, Office of Electricity for funding this work.





\bibliographystyle{IEEEtran}
\vspace{-1.5mm}
\bibliography{MBBconf_short}
%



\end{document}